**Multi-Valued Logic Gates based on Ballistic Transport in Quantum Point Contacts**

M. Seo[1], C. Hong[1], S. -Y. Lee[1], H. K. Choi[3], N. Kim[2], Y. Chung[1], V. Umansky[3] and D. Mahalu[3]

[1]*Department of Physics, Pusan National University, Busan, 609-735, Republic of Korea*

[2]*Korea Research Institute of Standard and Science, Daejeon, 306-600, Republic of Korea*

[3]*Department of Condensed Matter Physics, Weizmann Institute of Science, Rehovot, 76100, Israel*

Multi-valued logic gates, which can handle quaternary numbers as inputs, are developed by exploiting the ballistic transport properties of quantum point contacts in series. The principle of a logic gate that finds the minimum of two quaternary number inputs is demonstrated. The device is scalable to allow multiple inputs, which makes it possible to find the minimum of multiple inputs in a single gate operation. Also, the principle of a half-adder for quaternary number inputs is demonstrated. First, an adder that adds up two quaternary numbers and outputs the sum of inputs is demonstrated. Second, a device to express the sum of the adder into two quaternary digits [Carry (first digit) and Sum (second digit)] is demonstrated. All the logic gates presented in this paper can in principle be extended to allow decimal number inputs with high quality QPCs.

**Introduction**

The conductance of a quantum point contact (QPC) differs from that of a classical resistor due to the quantized energy levels and ballistic transport in the conduction channel. As a result, the output conductance of a QPC is quantized by 0, 1, 2, 3, 4… × $2e^2/h$ [1, 2], which provides an opportunity to develop multi-valued logic gates that can handle more than just binary numbers as an input. Another interesting property of a QPC is that the series resistance of multiple QPCs is determined solely by the QPC with highest resistance; the resistance of the other QPCs is irrelevant [3, 4]. Such unconventional transport properties can be exploited to develop logic gates that are fundamentally different from conventional logic gates based on an inverter. This paper reports a logic gate that finds the minimum of two quaternary number inputs based on the quantum transport properties of QPCs in series. The logic gate presented in this paper is scalable to allow multiple inputs, which makes it possible to find the minimum of multiple inputs in a single gate operation. The principle of a half-adder, which finds the sum and carry of two quaternary numbers, is then demonstrated. Although the logic operations for quaternary number inputs were observed, all the logic gates presented in this paper can in principle be extended to allow decimal number inputs with high quality QPCs. The logic gates presented in this work are expected to be useful for enhancing the sorting algorithms and reducing the computation time significantly.

**Results**

The total resistance of classical resistors in series is given by the sum of all resistors. On the other hand, the resistance of QPCs in series is determined by the resistance of the QPC with the maximum resistance provided that the transport inside the QPCs is ballistic [3, 4]. All other QPCs with lower resistances are irrelevant. This can be expressed in terms of the conductance as $G_s = \min[G_1, G_2, ...., G_N]$, where $N$ is the number of QPCs, $G_n$ is the conductance of the individual QPC and $G_s$ is the conductance through the series QPCs, which is quantized by integer multiples of $G_0 = 2e^2/h$. Wharam et al. [3] reported that the series resistance of two QPCs in series behaves as predicted [4].

By exploiting this interesting phenomenon, a 'minimum logic gate' that finds the minimum value of inputs can be realized. In principle, the device with only two QPCs in series can be used as a minimum-finding logic gate (MFLG). In practice, however, it is necessary to decouple two QPCs electrostatically so that the number of conducting channels in the QPCs can be assigned independently without affecting the nearby QPC. A MFLG device, which employs three QPCs in series to decouple the two input QPCs electrostatically, was developed. The device was fabricated on a GaAs/AlGaAs heterostructure wafer (Fig. 1a and Method). Unlike the conventional QPCs, an extra middle gate (MG in the figure) was used in the gap of the QPC gates. By applying a positive voltage on the middle gate, the depth of the potential well in the QPC could be made deeper and sharper, thereby making the sub-band energy spacing a few times larger (~7 meV) than that of a conventional QPC [5]. This helps to observe clear conductance quantization at 4.2 K and fabricate a clean 1-dimensional conducting channel along the gate by minimizing the effects of unwanted potential fluctuations drawn by the ionized impurities in the modulation doping region.

Fig.1b shows the conductance through the MFLG device measured as a function of the voltage on the QPC1(Q1) and QPC2(Q2). A positive voltage (typically +0.25 V, just below the voltage that makes the current leak to the collector drains C1 and C2) on the decoupling gate DC was applied to decouple Q1 and Q2 electrostatically. The device was totally pinched in the dark-blue region, whereas several constant-color (blue, light blue, green, yellow etc.) regions showed quantized conductance (~1, 2, 3, 4×$2e^2/h$ etc.). The inset is the derivative of the conductance ($[(dG/dV_{Q1})^2+(dG/dV_{Q2})^2]^{1/2}$). The white vertical (horizontal) lines appear as the number of conducting channels in Q1 (Q2) changes by the Q1 (Q2) gate voltage $V_{Q1}$ ($V_{Q2}$). The transition lines between the plateaus were almost vertical and horizontal, which means that the two input QPCs barely affect each other. This allows the two inputs to be assigned independently. For comparison, Fig. 1c presents the same measurements for the device without a decoupling gate (two 100nm-wide QPCs in series separated by a

100nm gap), which is similar to the results reported by Wharam et al. [3]. The transition voltages were determined by both the Q1 and Q2 voltages, which makes it difficult to use it as a logic gate device.

The output conductance of 0, 1, 2, $3\times 2e^2/h$ can be considered as the output states 0, 1, 2 and 3, whereas the QPC gate voltages that give the corresponding output conductance can be considered as the input states 0, 1, 2 and 3 (shown as white dashed lines in Fig. 1b). Fig. 1d shows the output conductance of the device measured for various inputs of Q1 and Q2 (the inputs are changed as a function of time). The output conductance was plotted after subtracting the ~1.1 K$\Omega$ series resistance. The figure shows that the output conductance of the device is determined only by the QPC with a smaller conductance, whereas the QPC with a higher conductance becomes irrelevant. The output conductance representing the '2' and '1' output states, marked by the red and blue arrows, showed small conductance steps. The different series resistances (non-ballistic resistance) of each QPC are responsible for the conductance steps, because the output conductance is determined only by the QPC with a smaller conductance. The height of the conductance step was typically less than 10% of $2e^2/h$, which is small enough to distinguish between the two different output states. In principle, the minimum-finding logic gate can be used to determine the minimum values for decimal numbers provided that the quantized plateaus of the QPCs are distinguishable up to the 10$^{th}$ conductance plateau.

The device is fundamentally different from the conventional digital comparator, which has three output states presenting (A>B), (A<B) and (A=B) not the minimum value itself. Having the minimum value as an output is a great advantage. For example, the number of input gates can be extended easily by cascading the MFLG into a tree-like structure because the output of the MFLG can be coupled directly to the input of the next stage MFLG. Another way of extending the number of input gates is simply to add more input gates in series provided that the ballistic transport is maintained in the device. Such a scheme will allow the minimum value among $n$ inputs to be found by a single gate operation. In contrast, the conventional algorithm, which uses a conventional digital comparator, requires at least $n$-1 gate operations. In addition, a conventional comparator can compare only binary numbers, whereas the MFLG can compare the decimal numbers in principle. This will allow significant enhancements of the sorting algorithms.

Decoupling two QPCs in series electrostatically make it possible to use the device as a MFLG. In the other extreme limit, where two QPCs are totally coupled electrostatically, the device can be used to add two multi-valued inputs and generate the sum as an output. Fig. 2a shows the device, which can be used as an adder for two quaternary numbers. Two QPCs Q1 and Q2 (blue and red) composed of two diagonally facing gates were used as the input gates of the adder. This makes both QPCs form constrictions almost in the same area (marked by a dashed circle). Fig. 2b shows the output conductance (inset) and its derivative as a function of the Q1 and Q2 gate voltages. The output conductance is changed almost identically by Q1 and Q2. The white dashed lines were assigned to represent the input states of the quaternary numbers. The output of the device was measured for various input states of Q1 and Q2, as shown in Fig. 2c. The output state is just a sum of the input states, Q1 and Q2. This suggests that the confinement potential in the constriction is determined simply by the sum of the Q1 and Q2 gate voltages. A few different combinations of gates as Q1 and Q2 were attempted and the best result was obtained for the combination shown in Fig. 2a.

Adding two multi-digit numbers requires more than just creating the sum of two single-digit numbers. For example, the sum of two quaternary numbers ranges from 0 to 6, which cannot be expressed as a single quaternary digit. Therefore, it is necessary to express the sum into two quaternary digits [Carry (first digit) and Sum (second digit)], as summarized in the table in Fig 3d. The device shown in Fig. 3a was used to generate the Sum and Carry when the input ranged from 0 to 7 (7 is also considered for the case when there is a Carry from a lower digit.). A positive voltage was applied to the B2 gate so that the current can flow to the drain D2 while keeping the B1 gate closed. The bias voltage of -100 μV was applied to the source S, and Q1 was used as an input gate while fixing the Q2 voltage to allow only 4 conducting channels to pass through the QPC, as shown in Fig. 3b. The number of conducting channels passing through Q1 depends on the Q1 input gate voltage. When the number of conducting channels in Q1 is less than or equal to 4, it is expected that most of electrons passing through Q1 will also pass through Q2 provided that the transport is ballistic. For more than 4 channels in Q1, the electrons carried by the extra channels will be rejected by Q2 and either deflected towards the D2 drain or back reflected towards the source.

Fig. 3c shows the conductance measured at D1 and D2 for a range of Q1 input states. The input state of Q1 was assigned for the Q1 voltages, as shown as pink dashed lines in Fig. 3c. The conductance was plotted without compensating for the series resistance of the device. The conductance of the plateaus measured at D1 were smaller than the corresponding integer multiples of $2e^2/h$. The conductance of the plateaus was recovered to the expected values by compensating 2.5KΩ series resistance. The conductance plateaus measured at D2 were approximately 0.63, 1.3 and 1.87×$e^2/h$ with a smaller step at approximately 0.62×$e^2/h$. The reduction of the conductance step cannot be explained entirely by the series resistance of the device because the resistances of these plateaus (~ 41.1, 19.9 and 13.8 KΩ) were comparably larger than the series resistance of the device (~ 2KΩ). A possible explanation for this reduction is that a considerable number of electrons are back-reflected towards the source than deflected towards the D2 drain, which is reasonable for a ballistic device. Nevertheless, all the plateaus were distinguishable and could be used to represent certain digital states.

The table in Fig. 3d shows the output characteristics of the device. The output states at D1 and D2 were different from the ideal sum and carry for the quaternary digit half-adder shown in the last two columns of the table. This discrepancy can be overcome by swapping the output of D1 and D2 using a conventional analog switch and comparator if the output state at D1 is equal or larger than 4, which is not so difficult. This will swap the partial columns shaded in blue and red in the table to match the expected Sum and Carry (A Carry of 4 rather than 1 can be corrected by dividing the signal with a simple resistor network). The overall operation will transform the input Q1, which ranges from 0 to 7, to two digit quaternary numbers. Cascading the device shown in Figs. 2a and 3a (The two devices can be merged into a single device by attaching the B1 and Q2 gates of Fig 3a at the output of the device shown in Fig. 2a) and swapping the partial output with an analog comparator and switch will produce a half-adder for quaternary numbers. In principle, the half-adder for the decimal input is also possible with high quality QPCs.

Discussion

In summary, this paper reports the principles of logic gates based on the ballistic transport properties of QPCs in series. All the logic gates presented in this paper can be enhanced to handle decimal numbers for input with high quality QPCs. Recently, a QPC showing distinguishable plateaus up to $20 \times 2e^2/h$ was reported [7].

The logic gates presented in this paper will help to reduce the size and complexity of the device needed to perform the same logic operation with conventional logic gates. For example, a conventional half-adder for binary numbers requires more than ten field-effect transistors (FET) at least. In contrast, the half-adder for multi-valued numbers presented in this paper requires only a few QPCs and a few FETs for an analog switch (4 FETs) and a comparator (1 FET).

The principles of the device rely on rather robust and simple physical phenomena, the ballistic transport and electrostatic effect, which increases the likelihood of its adaption to real applications. The only impediment is to increase the working temperature of the QPC up to room temperature. For the moment, the highest temperature to observe the conductance quantization in QPC is approximately 30K [8]. Although the principles of MFLG and half-adder were presented only, we believe that different logic gates can also be developed based on the ballistic transport properties of QPCs in series.

Methods

The devices were fabricated on a conventional uniform doped GaAs/GaAlAs heterostructure grown by molecular beam epitaxy. The 2-dimensional electron gas was buried 77 nm below the surface of the

GaAs/AlGaAs heterostructure. The carrier density and mobility were $2.9\times10^{11}$ cm$^{-2}$ and $2.5\times10^{6}$ cm$^{2}$/Vs, respectively, at 4.2 K.

The widths of the all the gates were 100 nm except for the 300 nm wide decoupling gate (DC). The gaps between the gates were kept to 100nm. The device was cooled to 4.2 K with a positive voltage (+0.5 V) on all gates to reduce the random telegraph noise of the device [6]. Therefore, the 2DEG beneath the gates were typically depleted below the gate voltage of approximately +0.3V. The middle gate voltage was set to approximately +0.75 V to open the conduction channels in the QPCs. The conductance was measured by applying a DC bias voltage of 100 μV to the source and measuring the drain current.

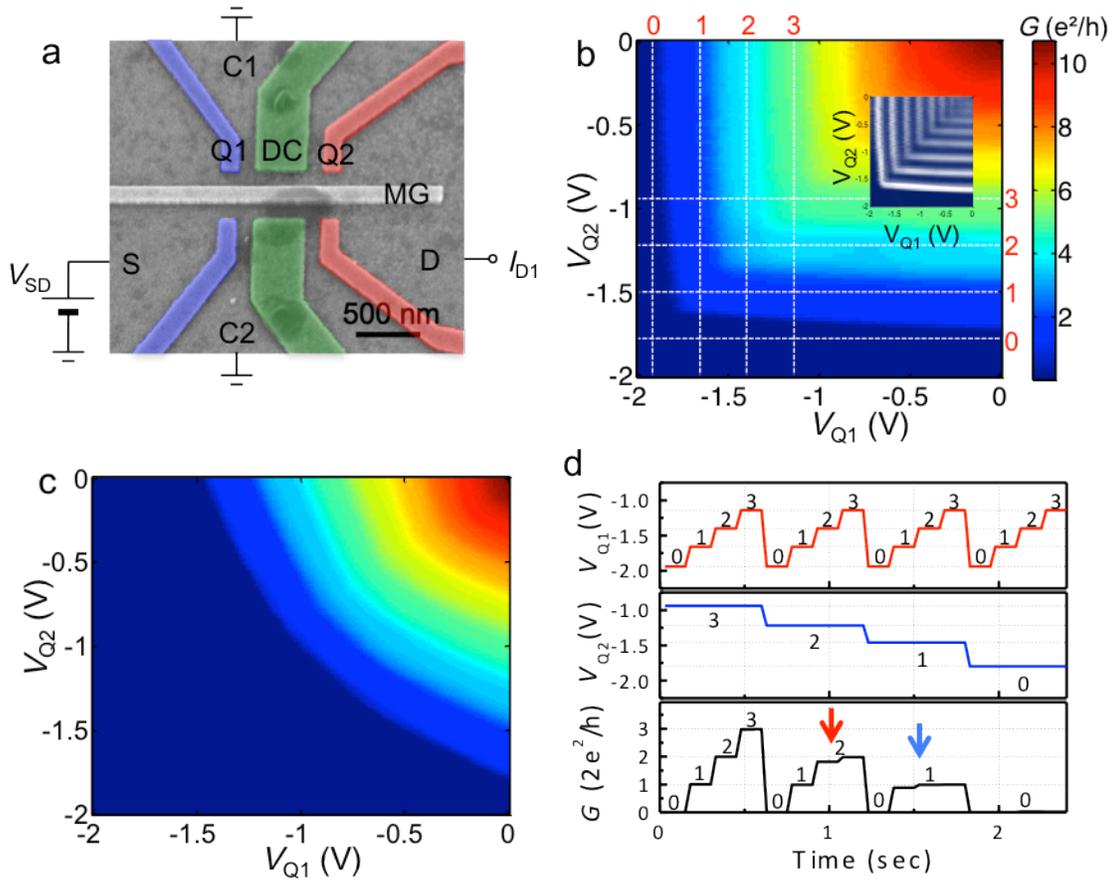

Fig. 1. **a,** Scanning electron microscopy (SEM) image of the minimum-finding logic gate device. A bias voltage was applied to the source S and the current through the device was measured from the drain D. Two collectors C1 and C2 were placed between the decoupling gates DC and grounded. **b,** Conductance $G$ measured through the device as a function of $V_{Q1}$ and $V_{Q2}$. The inset is the derivative of the conductance $[(dG/dV_{Q1})^2+(dG/dV_{Q2})^2]^{1/2}$. The derivative is larger for a whiter color, whereas the dark blue color represents zero derivatives. The white dashed lines show the voltages for Q1 and Q2 assigned for the 0, 1, 2 and 3 input states. **c,** Conductance measured for two QPCs in series without decoupling gate. **d,** Output state measured for a range of Q1 and Q2 input states. The output states were determined by the minimum input state of Q1 and Q2.

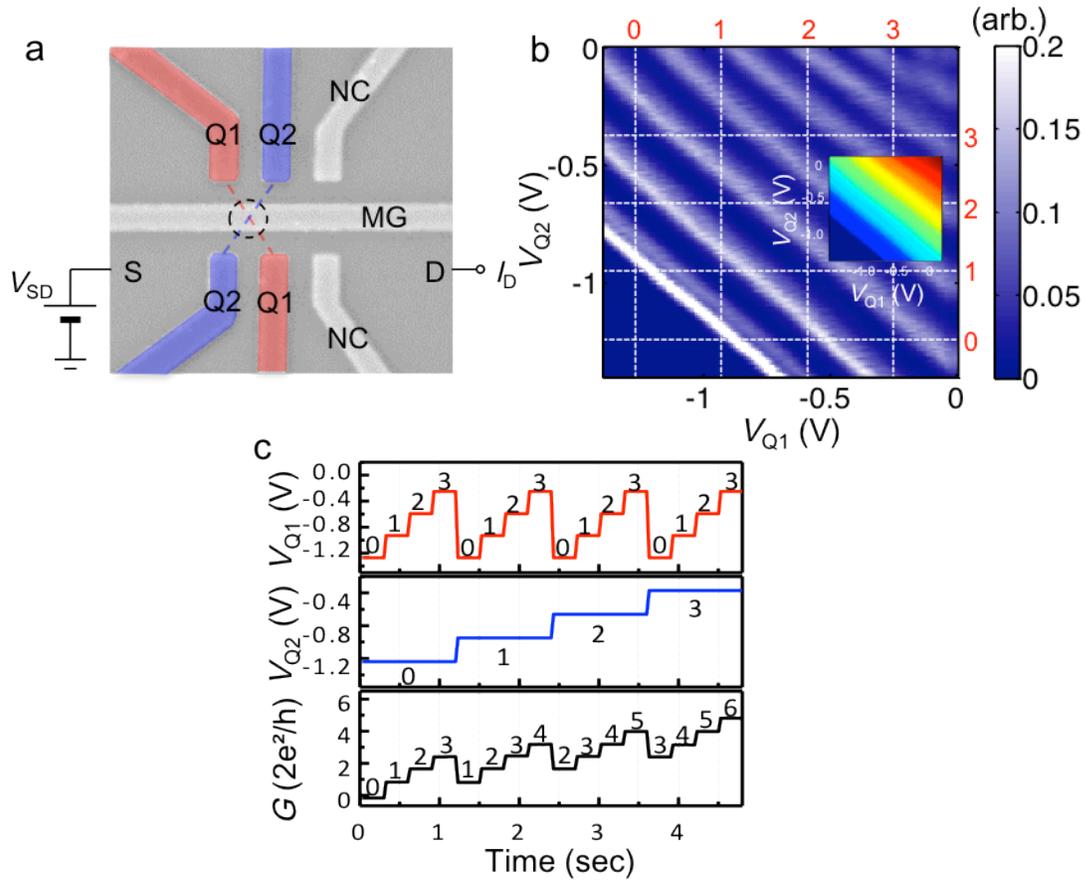

Fig. 2. **a,** Adder for multi-valued inputs. The diagonally facing gates (blue gates and red gates) were combined to form QPC Q1 and Q2. Extra gates (NC) on the right of the two QPCs were not used. **b,** Derivative (inset is the conductance) of the conductance $[(dG/dV_{Q1})^2+(dG/dV_{Q2})^2]^{1/2}$ measured through the device. The white dashed lines show the voltages for Q1 and Q2 assigned to the 0, 1, 2 and 3 input states. **c,** Output state measured for a range of Q1 and Q2 input states. The output state is the sum of the input states of Q1 and Q2.

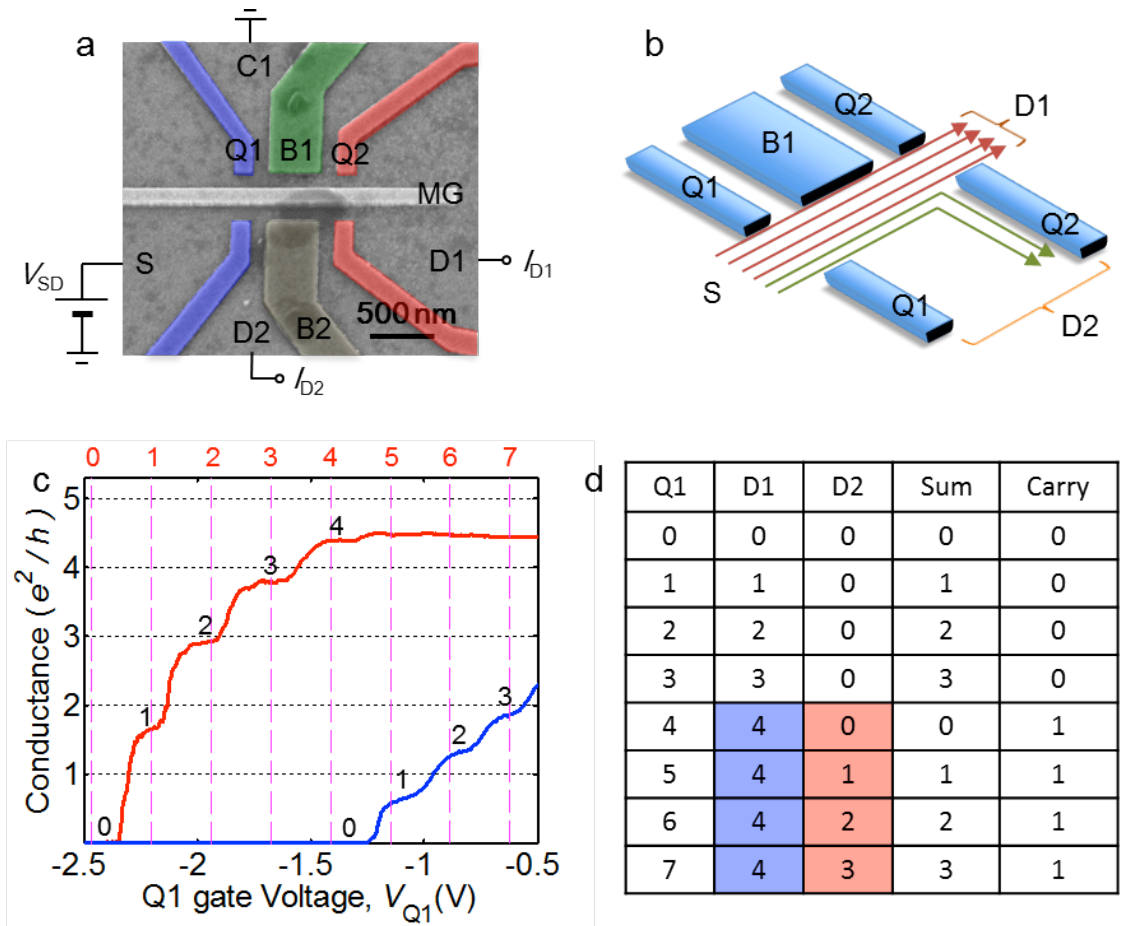

Fig. 3. **a,** Device to generate Sum and Carry for a given input which ranges form 0 to 7. The device is the same as the minimum-finding logic gate, except for the B2 gate is open to collect the current from D2. **b,** Illustration showing the concept of the device. Q2 was set to allow only 4 conducting channel to pass through. Extra channels, injected form Q1, will be deflected towards the drain D2. **c,** Conductance measured by varying the Q1 gate voltages. The Q1 gate voltages marked by the dashed red lines were assigned to the input states from 0 to 7. **d,** The results are summarized as a table. The last two columns are the proper Sum and Carry for the quaternary number half-adder.

**Acknowledgements**

This study was supported by the Basic Science Research Program through the National Research Foundation of Korea (NRF) funded by the Ministry of Education (NRF-2013R1A1A2010792). The authors also wish to thank M. Heiblum for the experimental support (access to his facilities and advice).

**Author Contributions**

M.S. and C.H. performed the experiments and equally contributed to this work. S.Y.L., N.K. and Y.C. planned the project. V.U. grew the wafers for the experiment and D.M. and H.K.C. assisted in fabrication. Y.C. supervised the project and prepared the manuscript. All authors discussed the results and commented on the manuscript.

**Additional Information**

Correspondence and requests for materials should be addressed to Y.C. (ycchung@pusan.ac.kr).

**Additional Information**

The authors declare no competing financial interests.